# Twist-phase-matching in two-dimensional materials

Hao Hong[1,2#]*, Chen Huang[1#], Chenjun Ma[1#], Jiajie Qi[1#], Can Liu[3], Shiwei Wu[4], Zhipei Sun[5], Enge Wang[6,7,8]*, and Kaihui Liu[1,6,7]*

[1] State Key Laboratory for Mesoscopic Physics, Frontiers Science Center for Nano-optoelectronics, School of Physics, Peking University, Beijing, China

[2] Interdisciplinary Institute of Light-Element Quantum Materials and Research Centre for Light-Element Advanced Materials, Peking University, Beijing, China

[3] Department of Physics, Renmin University of China, Beijing, China

[4] State Key Laboratory of Surface Physics and Department of Physics, Fudan University, Shanghai, China

[5] Department of Electronics and Nanoengineering and QTF Centre of Excellence, Aalto University, Aalto, Finland

[6] International Centre for Quantum Materials, Collaborative Innovation Centre of Quantum Matter, Peking University, Beijing, China

[7] Songshan Lake Materials Lab, Institute of Physics, Chinese Academy of Sciences, Dongguan, China

[8] School of Physics, Liaoning University, Shenyang, China

[#] These authors contributed equally to this work

* Correspondence: khliu@pku.edu.cn, haohong@pku.edu.cn, egwang@pku.edu.cn

**Abstract:**

Optical phase-matching involves establishing a proper phase relationship between the fundamental and generated waves to enable efficient optical parametric processes. It is typically achieved through either birefringence or periodically assigned polarization. Here, we report that twist angle in two-dimensional (2D) materials can generate a nonlinear Pancharatnam-Berry optical phase to compensate the phase mismatch in the process of nonlinear optical frequency conversion, and the vertical assembly of 2D layers with a proper twist sequence will generate a nontrivial "twist-phase-matching" (twist-PM) regime. The twist-PM model offers superior flexibility in the design of optical crystals, which works for twisted layers with either periodic or random thickness distribution. The designed crystals from twisted rhombohedra boron nitride films give rise to a second-harmonic generation conversion efficiency of ~8% within a thickness of only 3.2 μm, and a facile polarization controllability that is absent in conventional crystals (from linear to left-/right-handed circular/elliptical polarizations). Our methodology establishes a platform for the rational designing and atomic manufacturing of nonlinear optical crystals based on abundant 2D materials for various functionalities.

2 / 15

Nonlinear optical frequency conversion, which arises from the high order oscillations of dipoles within materials driven by a strong optical electric field, plays a crucial role in both classical and quantum optics[1-3]. This nonlinear process is rather weak and requires a long light-matter interaction path to achieve a considerable conversion efficiency. During the propagation of fundamental light and generation of nonlinear signals in the thick optical crystal, chromatic dispersion causes a difference in wavevectors, resulting in destructive interference between parametric waves from different parts of the crystal and a low conversion efficiency. To compensate this wavevector difference, birefringent-phase-matching (birefringent-PM) or quasi-phase-matching (quasi-PM) technique is employed to maintain a continuous energy flow from the fundamental to the parametric waves[4-10]. Birefringent-PM involves selecting an appropriate crystal direction that exploits birefringence to equalize the refractive indices of both waves. Quasi-PM is to add a periodic structure to induce an additional momentum to balance out the wavevector difference. Over the past few decades, many materials with high optical nonlinearities have been grown[11], and then fashioned into optical crystals, waveguides, microrings, and microdisks, guided by these two principles[12-15]. These advancements have significantly propelled the development of modern optics and photonics including quantum light sources, photonic integrated circuits, ultrafast and high-power lasers.

Two-dimensional (2D) atomic layered materials have long been supposed to be the potential revolutionary materials for nonlinear optical crystals[16-28], due to their ultrahigh nonlinear susceptibility[23], facile fabrication ability for integration[29], compatibility with Si-based optical chips[30], and rich material library for choosing and engineering[31]. Recently, great efforts have been devoted to fabricating thick 2D material crystals, e.g., $MoS_2$, $PdSe_2$, $NbOI_2$ and $NbOCl_2$, and bringing a quite strong second-harmonic generation (SHG) response[32-36]. However, the highest conversion efficiency has just approached 0.2% (ref. [34]), which is far below that of traditional bulk crystals[11], due to the absence of a suitable phase-matching design. The difficulty lies in the fact that conventional phase-matching techniques are not directly applicable to most kinds of 2D materials. Because atomic layered materials suffer from a tendency to cleave along the in-plane direction that constrains cutting along other directions in the birefringent-PM design, and their structural domain orientations are difficult to be poled by either electric, optical, or magnetic fields for quasi-PM engineering.

As a unique engineering method for 2D materials, interlayer twisting has brought emergent physics and functional applications, e.g., moiré excitons, superconductivity, ferroelectricity, and mid-infrared photodetectors[37-39]. This magic tool also provides a new degree of freedom, in principle, for intentionally manipulating the optical phase through stacking 2D layers. In this work, we introduced the twist angle into the nonlinear optical crystal design in the exampled material platform of 2D rhombohedra boron nitride (rBN). We discovered that twist angle can compensate phase mismatch in SHG process, and the vertical assembly of rBN films with proper interfilm twist angles will generate a nontrivial twist-PM regime. The boundaries of twist-PM are just quasi-PM and perfect-PM. Under the guidance of this model, we realized a SHG conversion efficiency of ~8% within a thickness of only 3.2 μm (the highest efficiency in all 2D family ever reported). Moreover, the generated output light polarization (i.e., linearly, circularly, or elliptically polarized) can also be directly controlled by the stacking orders, which has long been pursued but absent in conventional bulk crystals.

In our experiments and calculations, the fundamental beam incidents along the out-of-plane $z$ direction. For a 2D material with in-plane 3-fold rotation symmetry and mirror symmetry, the nonzero component of the second-order susceptibility is $\chi^{(2)} = \chi_{xxx} = -\chi_{xyy} = -\chi_{yyx} = -\chi_{yxy} = \chi$, where the armchair and zigzag directions lie along the $x$ and $y$ coordinates, respectively. The second-order polarization can thus be written as: $P_x = \varepsilon_0\chi(E_x^2 - E_y^2)$, $P_y = -2\varepsilon_0\chi E_x E_y$, or from view of the circular polarization:

$$\begin{cases} P_- = \sqrt{2}\varepsilon_0\chi E_+^2 \\ P_+ = \sqrt{2}\varepsilon_0\chi E_-^2 \end{cases} \quad (1)$$

where $\boldsymbol{P} = (P_-, P_+)^T = (P_x - iP_y, P_x + iP_y)^T/\sqrt{2}$, $\boldsymbol{E} = (E_-, E_+)^T = (E_x - iE_y, E_x + iE_y)^T/\sqrt{2}$).

Under a rotation operation $R = \begin{pmatrix} e^{i\theta} & 0 \\ 0 & e^{-i\theta} \end{pmatrix}$, the polarization evolves as:

$$\begin{cases} P_-' = \sqrt{2}\varepsilon_0 \chi E_+^2 e^{i3\theta} \\ P_+' = \sqrt{2}\varepsilon_0 \chi E_-^2 e^{-i3\theta} \end{cases} \quad (2)$$

It means that the rotation operation does not change the amplitude of the SHG polarization, but instead introduces a phase shift. In fact, this can be understood as the nonlinear Pancharatnam-Berry phase in the field of metamaterials and metasurface.

The $\pm 3\theta$ phase shift introduced at the interface can be appropriately employed to compensate the phase mismatch during SHG process. More concretely, a simple substitution from Eq. (2) into the nonlinear coupled-wave equations in the slowly varying amplitude approximation $\frac{\partial A_2}{\partial z} = \frac{i2\omega}{2\varepsilon_0 cn(2\omega)} P_{NL}(K_2, 2\omega) e^{i(2\omega t - k_2 z)}$ leads to the twist-angle dependent nonlinear wave generation (providing that the pump light is almost undepleted):

$$\begin{cases} E_{-,2\omega} = \frac{i\sqrt{2}\omega}{cn(2\omega)} \chi E_{+,\omega}^2 \int_0^T e^{i[3\theta(z) - \Delta k z]} dz \\ E_{+,2\omega} = \frac{i\sqrt{2}\omega}{cn(2\omega)} \chi E_{-,\omega}^2 \int_0^T e^{i[-3\theta(z) - \Delta k z]} dz, \end{cases} \quad (3)$$

where $E_{-,2\omega}$ ($E_{+,2\omega}$) and $E_{-,\omega}$ ($E_{+,\omega}$) are the left (right) circularly polarized components of the second harmonic and fundamental waves, respectively. $c$ is the speed of light in vacuum, $n(2\omega)$ is the refractive index of the second harmonic wave, $\Delta k = k_{2\omega} - 2k_\omega$ is the wavevector mismatch for SHG, and $\theta(z)$ is the twist angle of the layer at depth $z$ relative to the surface layer at $z = 0$ (fix $\theta(0) = 0$). It can be observed in Eq. (3) that a perfect twist-PM occurs when $\theta(z) = \Delta k z/3$ ($-\Delta k z/3$), with the incidence of right (left) circularly polarized fundamental light.

In our experiments, crystalline rBN films with a parallel-stacking interlayer configuration (Fig. 1(a) and in Supplemental Material [40]) are selected as the engineering platform for nonlinear optical crystal design, due to its large optical nonlinear susceptibility, broadband transparency, excellent physicochemical stability and high laser damage threshold[23,41-43]. The high-quality rBN films with thickness up to several micrometers are grown on the FeNi single crystal. The thickness dependent SHG intensity under 800 nm fundamental light excitation is measured in a transmission geometry. A quadratic dependence of the output signal on the thickness is expected when the film is thin (Fig. 1(d)), and then evolves to a sinusoidal dependence with increasing thickness (Fig. 1(e)), indicating the appearance of phase mismatch (Fig. 1(b)). The coherence length ($l_c$) is determined to be ~1.6 μm. To apply the twist-PM for efficient SHG process, we assembled two films with thickness of $l_c$, and monitored the evolution of SHG intensity with the twist angle between two films (Fig. 1(c)). A $\sin^2\frac{3\theta}{2}$ SHG intensity dependence is observed (Fig. 1(f)), due to a phase difference of $(\pi \pm 3\theta)$ between these two films. The output SHG intensity reaches a maximum at an interfilm twist angle of 60°, which corresponds to the conventional quasi-PM condition of periodic anti-parallel aligned polarizations (twist angle of 60° equals to 180° in rBN crystals).

In fact, the rotational freedom of twist angles facilitates a new designing regime far beyond conventional quasi-PM realized through switching electric polarizations. Based on the perfect-PM in twisted 2D films described in Eq. (3), one can freely design thick optical crystals from thin films. For example, we consider a quite simple stacking configuration consisting of $N$ films with an identical thickness of $t$ and twist angle of $\theta_m = (m-1)\theta$ for the $m^{\text{th}}$ film (Fig. 2(a)). The output SHG signal under circularly polarized light excitation is predicted to be

$$E_{\mp,2\omega} = \frac{i\sqrt{2}\omega}{cn(2\omega)}\chi E_{\pm,\omega}^2 \text{sinc}\left(\frac{\Delta k t}{2}\right)\frac{\sin\frac{N}{2}(\pm 3\theta - \Delta k t)}{\sin\frac{1}{2}(\pm 3\theta - \Delta k t)} t \cdot e^{i\frac{N-1}{2}(\pm 3\theta - \Delta k t) - i\frac{\Delta k t}{2}}, \quad (4)$$

The output SHG intensity with different parameters $(t, \theta)$ under a fixed total thickness is simulated and shown in Fig. 2(c). Twist-PM can be achieved under the condition of:

$$\theta = \pm \Delta k t /3, \quad (5)$$

The physical picture of twist-PM can be understood as illustrated in Fig. 2(b). Each layer corresponds to a very small vector which can be seen as a part of a radius-fixed circle, and each film contributes an arc chord vector, whose magnitude is determined by the film's thickness. The sum of the arc chord vectors corresponds to the electric field of SHG response. Twist-PM is to rotate the vectors from different segments into the same direction for a maximum output by a sequence of proper interfilm twist angles.

It is apparent that twist-PM represents a wide phase-matching regime for SHG, boundaries of which yield the perfect-PM ($\theta \to 0°$, $t \to 0$) and quasi-PM ($\theta = 60°$, $t = l_c$) (Fig. 2(d)). Naturally, the stacking of $\theta = 0°$ will lead to the phase-mismatching condition. These predictions are highly consistent with our experiments, where twisted rBN crystals with $\theta = 60°$ and $\theta = 0°$ stacking of 5 films (each with thickness of $l_c$) provide coherently enhanced SHG and sinusoidal oscillations, respectively (Fig. 2(d), dots). Obviously, the SHG efficiency in twist-PM regime increases with decreasing $t$, and approaches the maximum at $t \to 0$ (perfect-PM, Fig. 2(e), curve). Experimentally, we designed a total thickness of $2l_c$ under two conditions, i.e., $t = l_c$ (two films, $\theta = 60°$, quasi-PM) and $t = l_c/2$ (four films, $\theta = 30°$, twist-PM). As predicted by our model, the SHG intensity of the twist-PM crystal was about two times stronger than that of the quasi-PM crystal (Fig. 2(e), dots). As a rBN film of coherence thickness (~1.6 μm) yields a conversion efficiency of 1% (see details in Supplemental Material [40]), the conversion efficiency is expected to be 4% by quasi-PM, while it will be promoted to ~8% in the twisted rBN crystal with total thickness of ~3.2 μm within the scenario of twist-PM.

Different from quasi-PM where up and down electric polarizations must be in a certain sequence, the twist-PM works for films with a random thickness distribution (Fig. 3(a)). The arc chord vectors from different thicknesses of rBN films can also be aligned with a series of appropriate twist angles. The output SHG electric field can be described as:

$$E_{\mp,2\omega} = \frac{i\sqrt{2}\omega}{cn(2\omega)}\chi E_{\pm,\omega}{}^2 e^{-i\frac{\Delta k t_1}{2}} \times \sum_{m=1}^{N} \exp\left\{i\left[\pm 3\theta_m - \Delta k\left(\sum_{n=1}^{m} t_n - \frac{1}{2}t_m - \frac{1}{2}t_1\right)\right]\right\} t_m \text{sinc}(\Delta k t_m/2), (6)$$

And the twist-PM condition for films with arbitrary thicknesses is:

$$\theta_m = \pm \Delta k \left( \sum_{n=1}^{m} t_n - \frac{1}{2} t_m - \frac{1}{2} t_1 \right) / 3, \qquad (7)$$

where $\theta_m$ is the $m^{\text{th}}$ film's twist angle relative to the first incident film, $t_m$ is the thickness of the $m^{\text{th}}$ film. This thickness-free choice makes it quite facile and practical to fabricate nonlinear optical crystals from 2D films. Experimentally, we choose four films with thickness of ~800, 600, 400 and 300 nm, respectively. Stacking the film with a sequence of (0°, 25°, 42°, 55°) can promise the output SHG to increase monotonously (Fig. 3(b), dots), as predicted by our twist-PM model (Fig. 3(b), curve).

Finally, we demonstrated that twist-PM can not only tune the output intensity but also control the output polarization, which is an appealing advantage that conventional bulk crystals do not have. Obviously, from Eq. (3) the relative phase and amplitude of the two polarizations can be tuned via stacking configuration. A simple prediction is that two pieces of films with $t = l_c/2$ and $\theta = 30°$ will yield a nearly perfect circularly polarized SHG output (Fig. 4a), which can be confirmed by our simulation (Fig. 4(b)) and experiment (Fig. 4(c)).

In summary, our study has presented an effective methodology for rational designing and atomic manufacturing of nonlinear optical crystals in the platform of 2D materials by employing twist-PM model. This model holds a wide regime of phase-matching condition and is highly flexible for nonlinear optical crystal design, as it works for in-plane isotropic materials with arbitrary stacking thicknesses. This powerful designing ability in twist-PM model, together with rich species, emergent physical properties and facile property tunability in twisted 2D materials, will bring us a new family of nonlinear optical crystals of stacked 2D materials for advanced applications, such as in compact parametric amplifiers, quantum photonics, on-chip laser sources and optical modulators.

# References

[1]  N. Bloembergen, *Nonlinear optics* (World Scientific, 1992).

[2]  Y. R. Shen, *The principles of nonlinear optics* (Wiley-VCH, 2003).

[3]  R. W. Boyd, *Nonlinear optics* (Academic press, 2008).

[4]  S. Zhu, Y. Y. Zhu, and N. B. Ming, Science **278**, 843 (1997).

[5]  S. N. Zhu, Y. Y. Zhu, Y. Q. Qin, H. F. Wang, C. Z. Ge, and N. B. Ming, Phys. Rev. Lett. **78**, 2752 (1997).

[6]  I. Dolev, I. Kaminer, A. Shapira, M. Segev, and A. Arie, Phys. Rev. Lett. **108**, 113903 (2012).

[7]  Y. P. Chen, W. R. Dang, Y. L. Zheng, X. F. Chen, and X. W. Deng, Opt. Lett. **38**, 2298 (2013).

[8]  B. Q. Chen, C. Zhang, C. Y. Hu, R. J. Liu, and Z. Y. Li, Phys. Rev. Lett. **115**, 083902 (2015).

[9]  Z. Cui, D. Liu, J. Miao, A. Yang, and J. Zhu, Phys. Rev. Lett. **118**, 043901 (2017).

[10] J. Lin, N. Yao, Z. Hao, J. Zhang, W. Mao, M. Wang, W. Chu, R. Wu, Z. Fang, L. Qiao, W. Fang, F. Bo, and Y. Cheng, Phys. Rev. Lett. **122**, 173903 (2019).

[11] G. G. Gurzadian, V. G. Dmitriev, and D. N. Nikogosian, *Handbook of nonlinear optical crystals* (Springer, 1999).

[12] J. Wang, F. Bo, S. Wan, W. Li, F. Gao, J. Li, G. Zhang, and J. Xu, Opt. Express **23**, 23072 (2015).

[13] J. Lin, Y. Xu, J. Ni, M. Wang, Z. Fang, L. Qiao, W. Fang, and Y. Cheng, Physical Review Applied **6**, 014002 (2016).

[14] R. Luo, H. Jiang, S. Rogers, H. Liang, Y. He, and Q. Lin, Opt. Express **25**, 24531 (2017).

[15] R. Wolf, Y. Jia, S. Bonaus, C. S. Werner, S. J. Herr, I. Breunig, K. Buse, and H. Zappe, Optica **5**, 872 (2018).

[16] Y. L. Li, Y. Rao, K. F. Mak, Y. M. You, S. Y. Wang, C. R. Dean, and T. F. Heinz, Nano Lett. **13**, 3329 (2013).

[17] C. J. Kim, L. Brown, M. W. Graham, R. Hovden, R. W. Havener, P. L. McEuen, D. A. Muller, and J. Park, Nano Lett. **13**, 5660 (2013).

[18] J.-L. Cheng, N. Vermeulen, and J. Sipe, New J. Phys. **16**, 053014 (2014).

[19] K. L. Seyler, J. R. Schaibley, P. Gong, P. Rivera, A. M. Jones, S. F. Wu, J. Q. Yan, D. G. Mandrus, W. Yao, and X. D. Xu, Nat. Nanotechnol. **10**, 407 (2015).

[20] H. Z. Liu, Y. L. Li, Y. S. You, S. Ghimire, T. F. Heinz, and D. A. Reis, Nat. Phys. **13**, 262 (2017).

[21] N. Yoshikawa, T. Tamaya, and K. Tanaka, Science **356**, 736 (2017).

[22] J. D. Cox, A. Marini, and F. J. G. de Abajo, Nat. Commun. **8**, 14380 (2017).

[23] A. Autere, H. Jussila, Y. Y. Dai, Y. D. Wang, H. Lipsanen, and Z. P. Sun, Adv. Mater. **30**, 1705963 (2018).

[24] T. Jiang, D. Huang, J. L. Cheng, X. D. Fan, Z. H. Zhang, Y. W. Shan, Y. F. Yi, Y. Y. Dai, L. Shi, K. H. Liu,

C. G. Zeng, J. Zi, J. E. Sipe, Y. R. Shen, W. T. Liu, and S. W. Wu, Nat. Photonics **12**, 634 (2018).

[25] G. Soavi, G. Wang, H. Rostami, D. G. Purdie, D. De Fazio, T. Ma, B. R. Luo, J. J. Wang, A. K. Ott, D. Yoon, S. A. Bourelle, J. E. Muench, I. Goykhman, S. Dal Conte, M. Celebrano, A. Tomadin, M. Polini, G. Cerullo, and A. C. Ferrari, Nat. Nanotechnol. **13**, 583 (2018).

[26] J. D. Caldwell, I. Aharonovich, G. Cassabois, J. H. Edgar, B. Gil, and D. N. Basov, Nat. Rev. Mater. **4**, 552 (2019).

[27] H. Hong, C. C. Wu, Z. X. Zhao, Y. G. Zuo, J. H. Wang, C. Liu, J. Zhang, F. F. Wang, J. G. Feng, H. B. Shen, J. B. Yin, Y. C. Wu, Y. Zhao, K. H. Liu, P. Gao, S. Meng, S. W. Wu, Z. P. Sun, K. H. Liu, and J. Xiong, Nat. Photonics **15**, 510 (2021).

[28] K. Khaliji, L. Martín-Moreno, P. Avouris, S.-H. Oh, and T. Low, Phys. Rev. Lett. **128**, 193902 (2022).

[29] F. Liu, W. J. Wu, Y. S. Bai, S. H. Chae, Q. Y. Li, J. Wang, J. Hone, and X. Y. Zhu, Science **367**, 903 (2020).

[30] D. Akinwande, C. Huyghebaert, C. H. Wang, M. I. Serna, S. Goossens, L. J. Li, H. S. P. Wong, and F. H. L. Koppens, Nature **573**, 507 (2019).

[31] N. Mounet, M. Gibertini, P. Schwaller, D. Campi, A. Merkys, A. Marrazzo, T. Sohier, I. E. Castelli, A. Cepellotti, G. Pizzi, and N. Marzari, Nat. Nanotechnol. **13**, 246 (2018).

[32] M. Zhao, Z. L. Ye, R. Suzuki, Y. Ye, H. Y. Zhu, J. Xiao, Y. Wang, Y. Iwasa, and X. Zhang, Light: Sci. Appl. **5**, e16131 (2016).

[33] J. Yu, X. F. Kuang, J. Z. Li, J. H. Zhong, C. Zeng, L. K. Cao, Z. W. Liu, Z. X. S. Zeng, Z. Y. Luo, T. C. He, A. L. Pan, and Y. P. Liu, Nat. Commun. **12**, 1083 (2021).

[34] I. Abdelwahab, B. Tilmann, Y. Z. Wu, D. Giovanni, I. Verzhbitskiy, M. L. Zhu, R. Berte, F. Y. Xuan, L. D. Menezes, G. Eda, T. C. Sum, S. Y. Quek, S. A. Maier, and K. P. Loh, Nat. Photonics **16**, 644 (2022).

[35] X. Y. Xu, C. Trovatello, F. Mooshammer, Y. M. Shao, S. Zhang, K. Y. Yao, D. N. Basov, G. Cerullo, and P. J. Schuck, Nat. Photonics **16**, 698 (2022).

[36] Q. Guo, X. Z. Qi, L. Zhang, M. Gao, S. Hu, W. Zhou, W. Zang, X. Zhao, J. Wang, B. Yan, M. Xu, Y. K. Wu, G. Eda, Z. Xiao, S. A. Yang, H. Gou, Y. P. Feng, G. C. Guo, W. Zhou, X. F. Ren, C. W. Qiu, S. J. Pennycook, and A. T. S. Wee, Nature **613**, 53 (2023).

[37] N. P. Wilson, W. Yao, J. Shan, and X. Xu, Nature **599**, 383 (2021).

[38] K. F. Mak and J. Shan, Nat. Nanotechnol. **17**, 686 (2022).

[39] D. Huang, J. Choi, C.-K. Shih, and X. Li, Nat. Nanotechnol. **17**, 227 (2022).

[40] See Supplemental Material for additional experiment details

[41] Y. Kubota, K. Watanabe, O. Tsuda, and T. Taniguchi, Science **317**, 932 (2007).

[42] G. Cassabois, P. Valvin, and B. Gil, Nat. Photonics **10**, 262 (2016).

[43] N. Tancogne-Dejean and A. Rubio, Sci. Adv. **4**, eaao5207 (2018).

**Acknowledgements:** This work was supported by the National Key R&D Program of China (2022YFA1403504, 2022YFA1403500, 2021YFB3200303, 2021YFA1400201 and 2021YFA1400502), National Natural Science Foundation of China (52025023, 51991342, 52021006, 92163206, 11888101, T2188101, 12104018), Guangdong Major Project of Basic and Applied Basic Research (2021B0301030002), the Strategic Priority Research Program of Chinese Academy of Sciences (XDB33000000).

**Figures and captions**

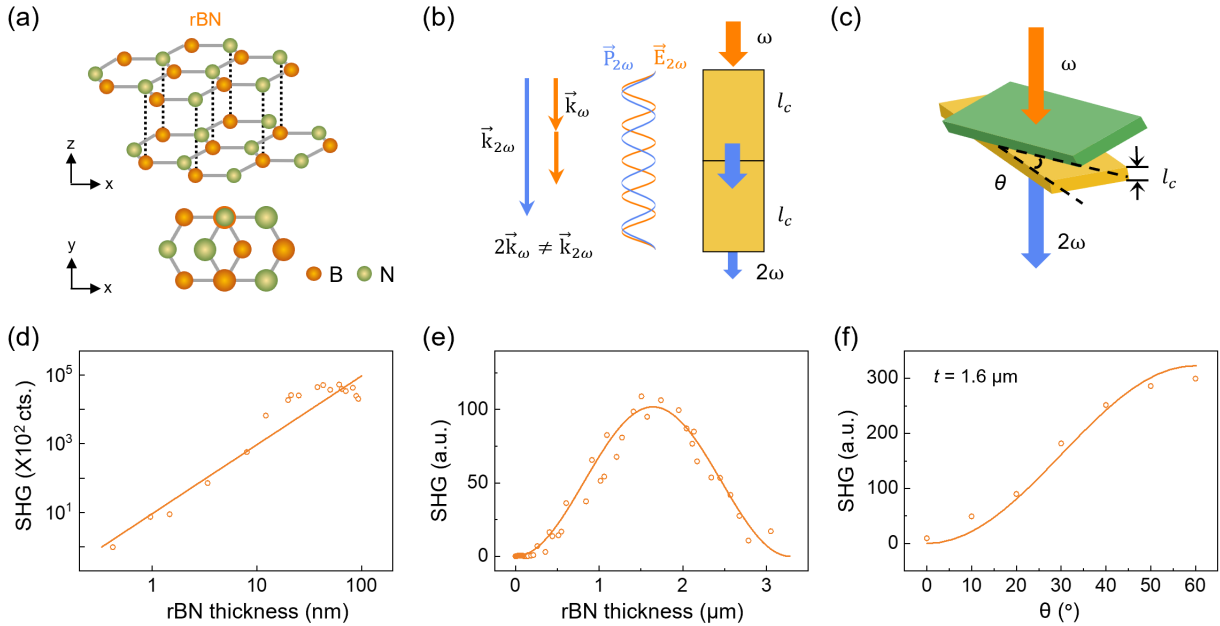

**FIG. 1 | SHG and phase mismatch in rBN films. (a)**, Schematics of the atomic structure of rBN. Each atomic layer is stacked in parallel to have the same polar orientation and bear an in-phase nonlinear optical response. **(b)**, Illustration of the phase mismatch during the propagation of fundamental wave and the generation of second harmonic wave in a nonlinear optical crystal. The chromatic dispersion causes a difference in wavevector ($\Delta k = k_{2\omega} - 2k_\omega$), resulting in destructive interference of SHG. **(c)**, Schematic of a nonlinear optical crystal assembled by two pieces of rBN films with an identical thickness of $t$ and twist angle of $\theta$. **(d, e),** Thickness-dependent SHG intensities. A quadratic dependence in **(d)** evolves to a sinusoidal dependence in **(e),** indicating a coherence length of $l_c \sim 1.6$ μm under 800 nm excitation. **f**, Twist angle dependent SHG output in twisted rBN sketched in **(c)**. The SHG intensity follows a $\sin^2 \frac{3\theta}{2}$ dependence on $\theta$, as demonstrated by our experiments (hollow circles) and theory predication (solid line).

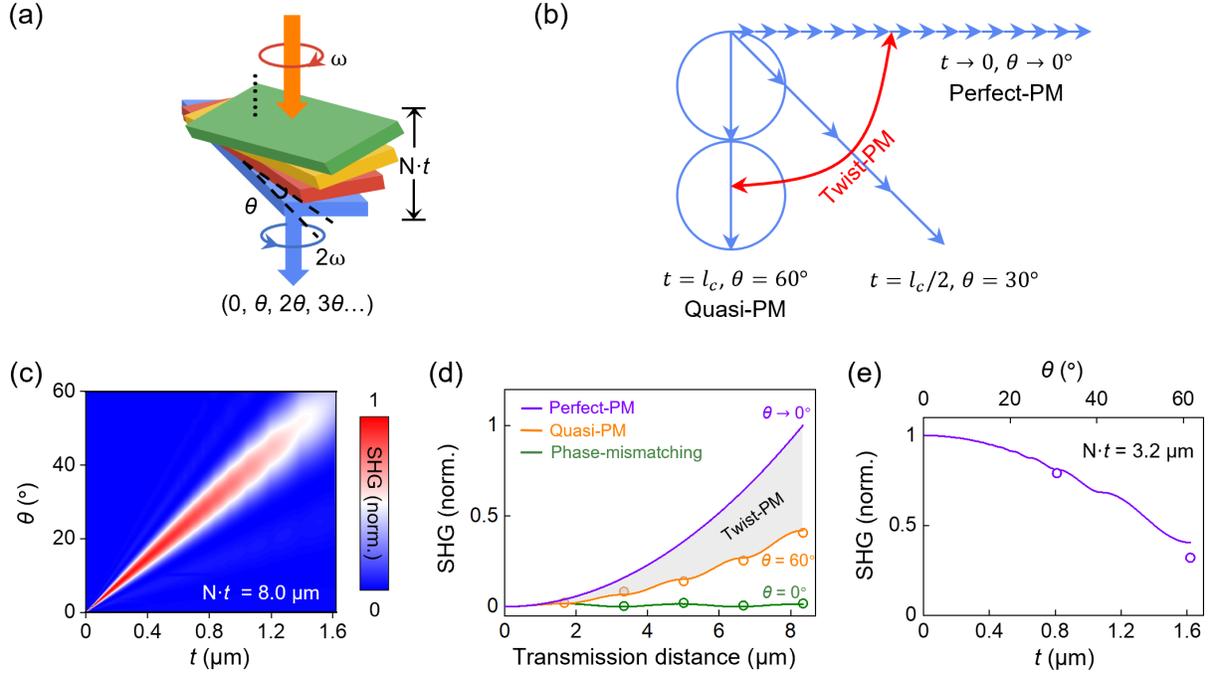

**FIG. 2 | Twist-PM for efficient SHG in twisted rBN films with the identical thickness. (a)**, Schematics of the realization of twist-PM through assembly of $N$ pieces of rBN films with an identical thickness of $t$ and incremental twist angle sequence. **(b)**, Illustration of physical picture of twist-PM process. The electric field of SHG is indicated as the arc chord vector, whose magnitude is determined by the film's thickness. Twist-PM is to align the arc chord vectors from different rBN film for a strong SHG response by the introduction of a proper optical phase at the interfilm. The boundaries of the twist-PM regime are just quasi-PM and perfect-PM. **(c)**, Simulated SHG intensity of twisted rBN sketched in **(a)** as a function of $t$ and $\theta$, with a fixed total thickness of 8 μm. Under the relationship of $\theta = \Delta k t / 3$, a twist-PM condition can be achieved for high-efficiency SHG. **(d)**, Theoretical (solid curves) and experimental (dots) results under different conditions of twisted rBN in **(a)**. The grey area denotes the twist-PM region, with the boundaries of perfect-PM ($\theta \to 0°$, violet curve) and quasi-PM ($\theta = 60°$, origin curve). The green curve represents the phase-mismatching ($\theta = 0°$). **(e)**, Output SHG intensity as a function of $t$ and $\theta$ under twist-PM conditions. The total thickness is fixed at $2l_c = 3.2$ μm. Twisted crystals under two twist-PM conditions, i.e., $t = l_c$ ($\theta = 60°$) and $t = l_c/2$ ($\theta = 30°$), have been measured experimentally (dots) and compared to the theoretical simulations (solid curves).

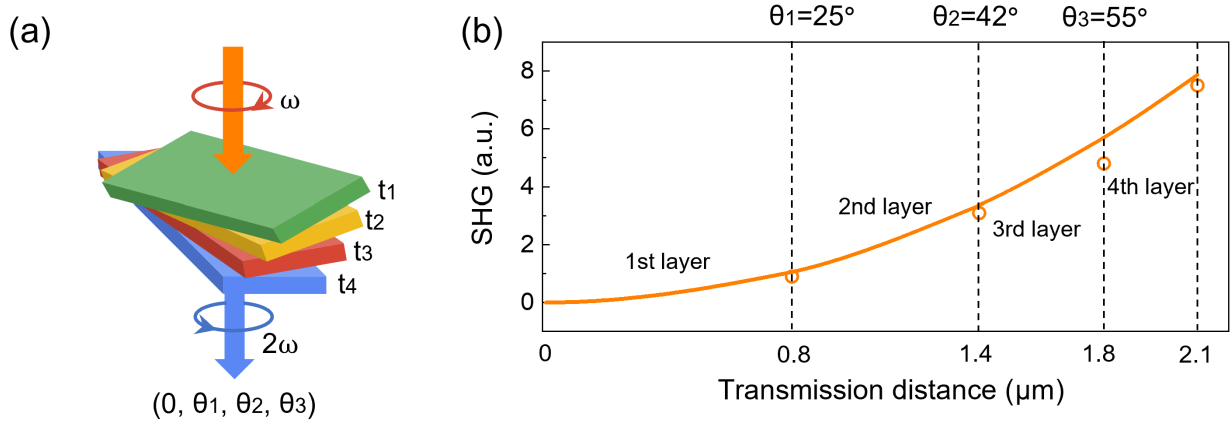

**Fig. 3 | Twist-PM for efficient SHG in twisted rBN films with random thicknesses. (a)**, Schematics of a nonlinear optical crystal assembled by four pieces of rBN films with different thicknesses of $t_1, t_2, t_3, t_4$, and a sequence of twist angles. **(b)**, Twist-PM for efficient SHG. For four rBN films with different thicknesses (800, 600, 400 and 300 nm), the twist-PM can still be fulfilled under twist angles of (0°, 25°, 42°, 55°), as demonstrated theoretically (solid line) and experimentally (hollow circles).

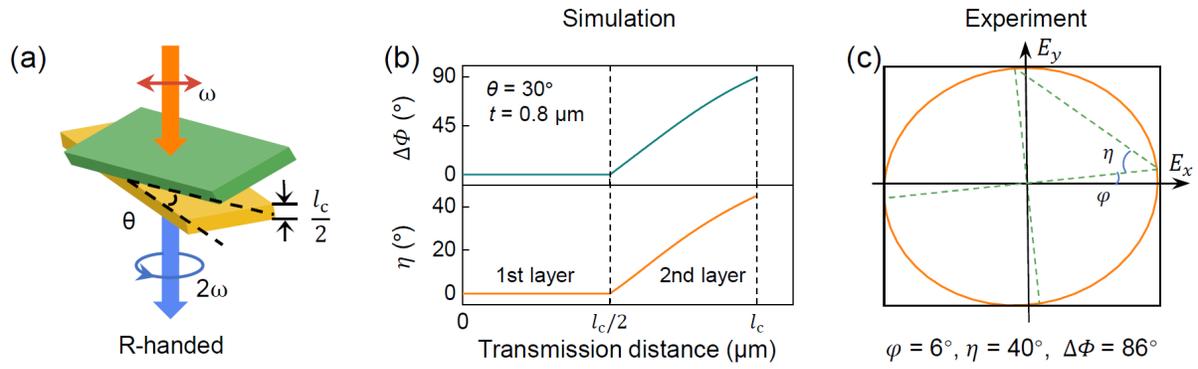

**Fig. 4 | Polarization control under the twist-PM model. (a),** Schematics of a nonlinear optical crystal assembled by two rBN films with an identical thickness $l_c/2$ and a twist angle $\theta = 30°$. **(b, c),** Simulation (**b**) and experiment (**c**) of the SHG polarization control in rBN films sketched in (**a**). Under linearly polarized light excitation, the SHG output can be tuned to be circularly.